# Characterization of a kerosene liquid jet injected in a high temperature Mach 2 supersonic crossflow

Authors : N. Fdida, N. Mallart-Martinez, T. Le Pichon, A. Vincent-Randonnier

Affiliation : DMPE, ONERA, Université Paris Saclay, F-91123 Palaiseau, France

Corresponding author: Nicolas.fdida@onera.fr

Abstract :

Near field liquid structures and penetration of a kerosene jet injected in a Mach 2 crossflow were studied experimentally in the LAPCAT-II Dual Mode Ramjet Combustor at Onera, using high spatial and temporal resolution imaging techniques. The experiments performed in this study provide measurements in high temperature conditions, with kerosene as liquid fuel to bring new data in these conditions. The fuel spray is injected through a single orifice, perpendicularly to the supersonic flow, at temperatures from ambient to 1500 K and jet-to-crossflow momentum flux ratio from 3.1 to 8.6. Five different test cases are defined to show independently the influence of the injection ratio and the supersonic flow temperature on the liquid jet atomization. A high spatial resolution imaging system is used to detail the characteristic spatial scales of this atomization process in the supersonic crossflow. The surface waves and droplets produced in the windward side of the liquid jet are characterized qualitatively. High-resolution visualizations bring a new insight of the droplet trajectories in the leeward region of the jet. Schlieren imaging is used to detail the complex shock wave structures in such a supersonic flow. Penetration of the kerosene jet is measured by shadowgraphy and schlieren imaging in five test cases and compared to other studies of the literature whenever it is possible. High-speed schlieren performed at 210 kHz is also performed to capture the temporal dynamics of the shocks and measure the liquid jet velocity.

List of symbols

| | |
|---|---|
| $d_j$ | Diameter of the fuel injection orifice (m) |
| $\dot{m}_j$ | Kerosene mass flow rate (g/s) |
| $\dot{m}_{air*}$ | Air free stream mass flow rate (g/s) |
| $\mu$ | Dynamic viscosity (Pa.s) |
| Ma | Mach number (dimensionless) |
| $P_0$ | Stagnation pressure (MPa) |
| $P_{inj}$ | Fuel injection pressure (MPa) |
| Re | Reynolds number (dimensionless) |
| q | Jet-to-crossflow momentum–flux ratio or injection ratio (–) |
| $T_0$ | Stagnation temperature (K) |
| $T_S$ | Static temperature |
| $u_{air*}$ | Air free stream velocity (m/s) |

| $u_j$ | Velocity of injected jet (m/s) |
|---|---|
| $We_{air*}$ | Gaseous Weber number (dimensionless) |
| $\rho_{air*}$ | Air free stream density (kg/m$^3$) |
| $\rho_j$ | Density of injected jet (kg/m$^3$) |
| $X_\alpha$ | Mass fraction of chemical species α |
| $Y$ | Penetration height (m) |
| $\nu$ | Kinematic viscosity (m²/s) |

# 1 Introduction

The air-breathing supersonic combustion ramjet (scramjet) is the most promising mode of propulsion for long-range hypersonic cruise vehicles due to its high specific impulse for flight Mach number ranging from 3 to 10, which means burning fuel with the atmospheric air instead of carrying both fuel and oxidizer (gain in volume and mass). Using scramjet engines is probably the most promising way to achieve this goal but many difficulties have to be addressed when designing such engines, such as improving ignition, stabilising the combustion in the supersonic internal flow or withstanding the high stagnation enthalpies with appropriate thermal protection systems. The way the fuel is injected in the combustor and its properties are part of the solutions for those two problems. Depending on the Mach of flight, supersonic combustion is controlled by the kinetics of combustion reactions, the mixing of the fuel with air in the combustor and the flame propagation. Up to Mach 7-8, liquid hydrocarbon fuels are considered as good candidates thanks to their high densities and their cooling capacities, especially if they undergo endothermic chemical reactions.

Transverse jets (using hydrogen or other hydrocarbon fuels) in supersonic crossflow have been widely investigated (see review from Sun et al., 2019), leading to canonical configurations for investigations of turbulent mixing and combustion at both the fundamental and the applied level (Gamba and Mungal, 2015). Whenever the fuel is injected in a liquid state, atomization and evaporation processes play a crucial role on the mixing delay, and have to be fast enough to allow combustion to take place in the combustor. Indeed, the coupling of atomization and vaporization processes with the supersonic hot air cross-flow leads to complex interactions of the liquid jet and induced shock wave structures. Compared to gaseous fuels, the liquid fuel atomization and vaporization plays a crucial role on mixing delay (Chang et al., 2018) and they drive the spatial distribution of the fuel droplets and consequently the flame topology. The deep penetration of atomized liquid fuel into the supersonic airflow promotes mixing and is important to sustain combustion (Ren et al., 2019). These processes then deserve to be investigated in depth. Many experimental studies can be found in the literature, dealing with sprays of water injected in a cold or hot supersonic flow (review by Ren et al., 2019) but open data are scarce in the case of liquid fuel injected in supersonic cross-flow of air at temperatures above 650 K. Moreover, the mechanisms of ignition and heat release on the mixing layer development have to be clarified. Consequently, a study has then been initiated at ONERA's LAERTE facility to characterize injection, atomization, mixing and vaporization of Jet-A1 fuel in a Mach = 2 cross-flow of vitiated air for stagnation temperatures ranging from 300 K to 1500 K.

Many injection strategies depending on the combustor configuration or the Mach flight number have been proposed to enhance atomization and vaporization, such as a single aerated liquid jet (Ghenai et al., 2009, K.C. Lin et al., 2020), tandem liquid jets (Sathiyamoorthy et al., 2020) or through an aeroramp (Anderson and Schetz, 2005). Moreover, injection configurations have been proposed to enhance the combustion stabilization with a cavity (Nakaya et al., 2005), in the wake of a pylon (Gruber et al., 2008) or with a strut injection (Chang et al., 2018). For the present study, the fuel is injected through a single orifice, perpendicularly to the supersonic flow, at different conditions of vitiated air temperature or fuel mass flow. In this configuration, atomization and mixing are the key processes to study, as they influence the shock waves and the vortices developing around the liquid jet. Downstream, they lead to the creation of droplets and their coalescence which directly impacts the combustion efficiency (Wei et al., 2020). Therefore it is important to capture the spatial and temporal dynamics of the spray to master the combustion in supersonic conditions. The atomization in such a supersonic flow results of physical processes attributed to primary and secondary breakups, but their relative importance in the case of the supersonic flow is still under discussion. The fuel spray characteristics can be obtained in these tough conditions by implementing experimental diagnostics to measure the penetration height of the fuel jet and the droplet properties, such as the droplet size and velocity. Many authors have used imaging systems such as high-speed shadowgraphy (Ghenai et al., 2009), schlieren methods (Beloki et al., 2009, Anderson and Schetz, 2005, Yu et al., 2005), PLIF (Wang et al. 2015) and more recently X-Ray imaging (Lin et al., 2020) to measure the penetration height of the spray and visualize the atomization process. Shadowgraphy and schlieren methods have the advantage to show the shock around the liquid jet. The droplet size and velocity distributions in supersonic flows can only be obtain with Phase Doppler systems, such as the PDPA (Phase Doppler Particle Analyzer, Lin et al., 2004) or the PDA (Phase Doppler Analyzer, Li et al., 2019). Both studies, performed with pure water jets in a supersonic crossflow, at Mach 1.94 and a maximal total temperature $T_0$= 533 K for Lin et al., 2004 and at Mach 2.84 and $T_0$= 300K for Li et al. 2019, show Sauter Mean Diameter (SMD) around 15 µm and 25 µm, respectively. Such small sizes are tough to measure accurately with high speed imaging systems. As a goal of this study is to investigate the effect of vitiated air temperature on the spray, droplets even smaller are expected. Consequently, a high spatial resolution imaging system has been set to detail the characteristic spatial scales of this atomization process. Moreover, performing optical diagnostics in such conditions is challenging due to the high velocity of the supersonic flow convecting the droplets at several hundreds of m/s (Lin et al., 2004). Thus, high-speed schlieren imaging is also used to resolve the temporal dynamics of the shocks and measure the liquid jet velocity. Indeed, atomization is a multiscale phenomenon which can only be revealed with different imaging systems operated at different spatial and temporal resolution, as proposed here.

After a brief description of the LAPCAT-II Dual Mode Ramjet Combustor and its operating conditions, the optical setups used to achieve high-speed or high-resolution shadowgraphy and high-speed schlieren and high-resolution shadowgraphy imaging will be described. The paper will then focus on the experimental results obtained in the stationary phase of the supersonic flow with the description of the atomization mechanism of the liquid jet in the supersonic stream. Based on shadowgraphy and schlieren images, the influence of the fuel massflow and the temperature of the supersonic air flow on the penetration depth of the fuel and its distance of vaporization have been evaluated for six different operating conditions. Velocity maps of the liquid jet were obtained close to the injection point thanks to cross-correlation methods developed at ONERA. Results are compared for the different test cases.

## 2 Experimental Setup

### 2.1 The ONERA LAERTE Facility and the LAPCAT-II Dual Mode Ramjet Combustor

Experiments are performed with the LAPCAT-II dual-mode ramjet combustor, at the ONERA-LAERTE facility, presented in Vincent-Randonnier et al. (2014). In order to work as a dual-mode ramjet, the kerosene fuelled combustion chamber has two parts and two injection stages. The first part is slightly diverging and is mainly dedicated to supersonic combustion at a high flight Mach number, whereas the second one allows subsonic combustion at a lower flight Mach number, with a thermal throat located near the chamber end (Scherrer et al., 2016). In the ramjet mode, the supersonic air flow enters the intake and slows down to subsonic through a series of shocks before entering the combustion chamber. In the scramjet mode, the flow remains supersonic through the combustor. The dual-mode scramjet bridges the gap between the ramjet and scramjet, using the same combustor for both combustion modes but operating with a thermal throat in the ramjet mode. This may theoretically enable a vehicle to be operated from flight Ma ≈ 3 to Ma≈8 with only minor engine modifications. Previous studies with experimental and computational comparisons of the LAPCAT-II supersonic combustor fuelled with hydrogen, provided a better knowledge to describe the mechanism underlying this reacting flow (Vincent-Randonnier et al.

2018). The present study is the first one to be performed with the LAPCAT-II combustor with injection of a liquid fuel (Jet-A1). Note that no means to ignite the combustion is employed. Only auto-ignition may occur.

The LAERTE facility with the LAPCAT-II combustor is presented in Fig. 1, as a schematic view of the facility (a), a picture of the combustor (b) and a schematic of its longitudinal cross-section with the locations of the side windows (c). The facility is operated in the blow-down mode with the combustor acting as a heat-sink. The combustor is fed in air, thanks to a 2-tons and 25 MPa storage, and heated by the two-stage hydrogen burner. The air entering the combustor is then vitiated with water vapour, the molar fraction of $O_2$ being kept at 0.21 by an upstream replenishment of oxygen (Fig. 1-a and Table 1).

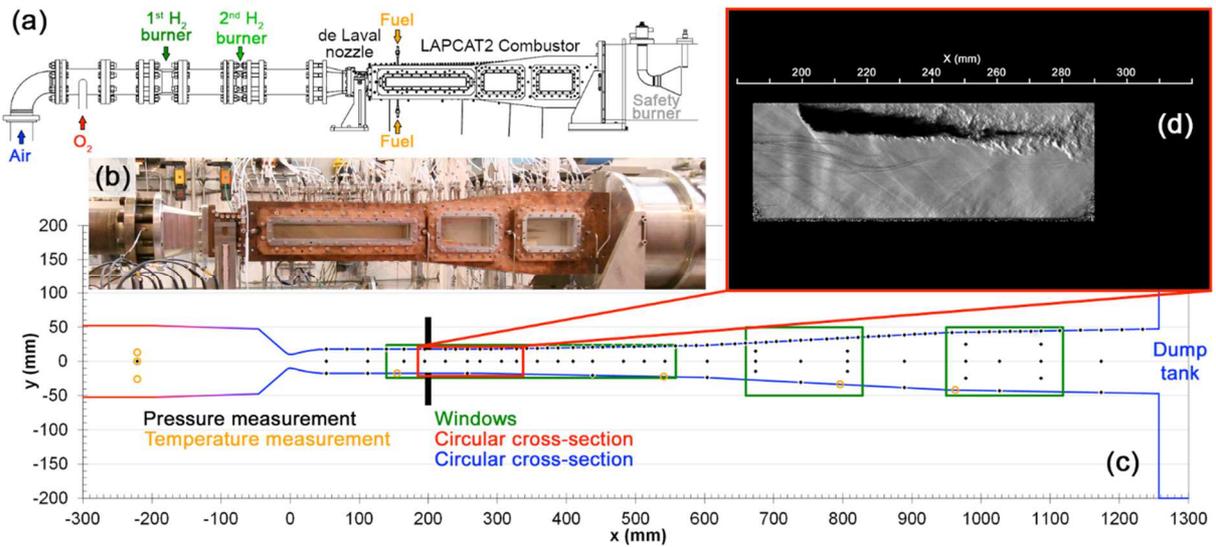

**Fig.1 (a)** Schematic plan of the LAERTE facility, **(b)** photo of the LAPCAT-II combustor, **(c)** geometry of the LAPCAT-II combustor **(d)** Instantaneous schlieren image in case HT-HQ, cf. Table 1

As indicated in Fig. 1c, the reference position, x= 0 mm, corresponds to the throat of the Mach 2.0 de Laval nozzle, used for this study. The combustor width is 40 mm and the inlet height is 35.4 mm. The de Laval nozzle and the combustion chamber are made from a copper alloy (CuCrZr) and the inner wall of which includes a 0.3 mm thick thermal barrier coating of Yttria stabilized Zirconia. The combustor is 1257 mm long and contains four successive sections: the first has a constant cross-sectional area, whereas the following sections have 1°, 3° and 1° of diverging half-angles respectively to prevent thermal choking. Windows are placed at different locations, allowing optical accesses for either single-point measurements or imaging techniques. The flow in the combustor is over-expanded and the flow transition from supersonic to subsonic occurs in the far downstream part of the combustor, more specifically in the third combustor section, around x≈750 mm to 800 mm, located in the second window.

Each run consists of a test sequence of ≈60 s providing 5 to 10 s of stationary injection conditions at the required $T_0$. During this test campaign, around hundreds of runs have been performed with experimental conditions tested for different $T_0$, vitiated air and fuel mass flows. A typical timing sequence of a test run is shown on Fig. 3, showing the temporal evolution of the stagnation temperature $T_0$ (light grey line), injection pressure (dashed line), static pressure $P_0$ (grey line) and mass flowrate of kerosene (black line). At t=15 s and at t=20 s the first and second preheating burners are successively activated, providing a maximum total temperature $T_0$ of 1500K (HT case). When only one preheating burner is activated, $T_0$ is around 600 K (MT case) and when there is no pre-burner in the sequence, $T_0$ is minimum at about 266K (LT case). The stationary phase of the supersonic flow begins when the temperature stabilizes at the requested $T_0$, typically around t0+27.5 s, when the final temperature is obtained (≈1500 K in Fig.2). The mean total temperature, with two pre-burners, over 36 runs is $T_{0-moy}$=1503 K +/-3K.

On Fig.2, the kerosene line is opened at t0=22 s, the liquid enters in the chamber at t=23.5 s and the jet mass flow rate $\dot{m}_j$ (g/s) is stationary from t0=24 s (depending on the flowrate) to t0=35 s, when the kerosene valve is closed and gaseous nitrogen purges the fuel line. This injection condition of Fig.2 corresponds to the maximum kerosene mass flow rate. For the minimum flow rate injection conditions ($\mathring{m}_j$=6.70 g/s), the stationary phase of the liquid mass flow is longer to establish and starts at t0+30 s (when 99% of the final mass flow rate is reached). The actual stationary phase of the kerosene liquid jet in the supersonic flow begins when both stationary phases of gaseous and liquid flow are reached. A typical instantaneous image of the kerosene jet injected in the test supersonic flow is illustrated on Fig.1d. Images are recorded from a trigger sent by the test sequence, at t0+20 s, at the opening of the kerosene valve. Measurements from shadowgraphs and schlieren imaging are made from images recorded during the stationary phase of the flow.

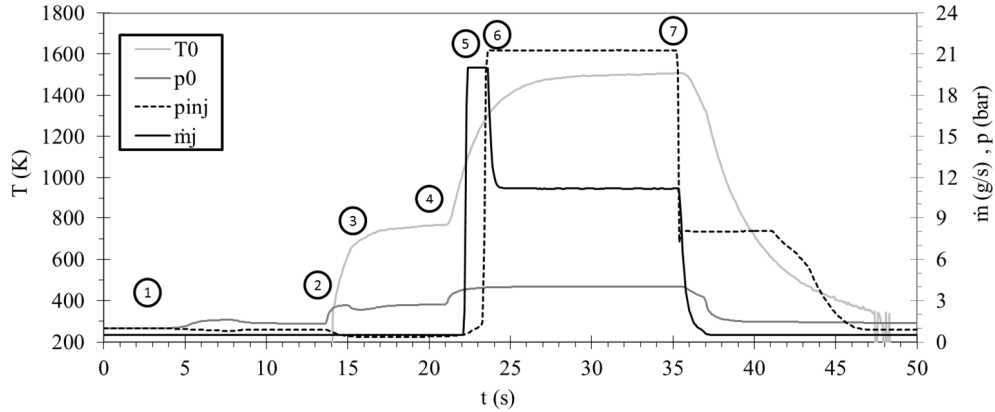

**Fig.2** Temporal evolution of the temperature, injection pressure and mass flowrate of kerosene during a HT-HQ test sequence. 1: air injection (t=2.5 s). 2: $H_2$ injection and ignition (heater, t=13 s). 3: $O_2$ injection (t=15 s). 4: $H_2$ injection ($2^{nd}$ heater, t=20 s). 5: fuel injection (t=22 s). 6: stabilized fuel injection (t=24 s). 7: fuel and $H_2$ injections stopped and $N_2$ flush valve opening (t=35 s)

Table 1 lists the different operating conditions and the associated characteristics of the air and fuel flows (respectively with g and j subscripts). The operating conditions have been chosen in order to study separately the influence of the liquid mass flow rate on the jet penetration and its atomization, as others research groups, and the influence of the temperature on the jet penetration, which has ever been done before in such supersonic conditions to the authors' knowledge. The operating conditions considered in this study focus on a supersonic combustion at a fixed total pressure of $P_0 \approx 0.40$ MPa, Ma=1.955 and three total temperatures $T_0$: 270 K and 610 K and 1500 K, respectively indicated as LT, MT and HT in Table 1. Kerosene Jet-A1 is injected at room temperature, at x=200 mm in the first visualisation section of the test rig as shown in Fig. 1.d. Three different flowrates are injected (LQ, MQ and HQ) through a single hole injector of theoretical diameter of 0.6 mm. The effective injector diameter $d_j$ determined experimentally from the linear relationship between the measured injection pressure and mass flow rate conduced to $d_j = 0.567$ mm. The effective diameter value corresponds to a discharge coefficient of about 0.77, which is consistent with Lubarsky et al., 2012, who investigated spray trajectories formed by Jet-A1 fuel injected into a vitiated air crossflow for different injector geometries. Indeed, in the case of a sharp edge orifice, similar to the injector orifice of this study, they determined a discharge coefficient of ≈0.75. Important non dimensional parameters used to figure out the forces acting on this flow are the gaseous Weber number defined as $We_{air*}=\rho_{air*} \cdot u_{air*}^2 \, d'_j/\sigma$, the Reynolds number based on the injector diameter $Re=u_j \cdot d'_j/\nu$ and the jet-to-crossflow momentum flux ratio q defined as $q=\rho_j u_j^2/\rho_{air*} u_{air*}^2$, where u is the flow velocity. The behaviour of the flow and combustion is characterized with 124 pressure measurements implemented along the combustor walls whereas the stagnation temperature $T_0$ is estimated based on 3 thermocouple measurements located at x=-220 mm. The data given in Table 1 are averaged values measured for each operating condition, during the stationary phase of the flow.

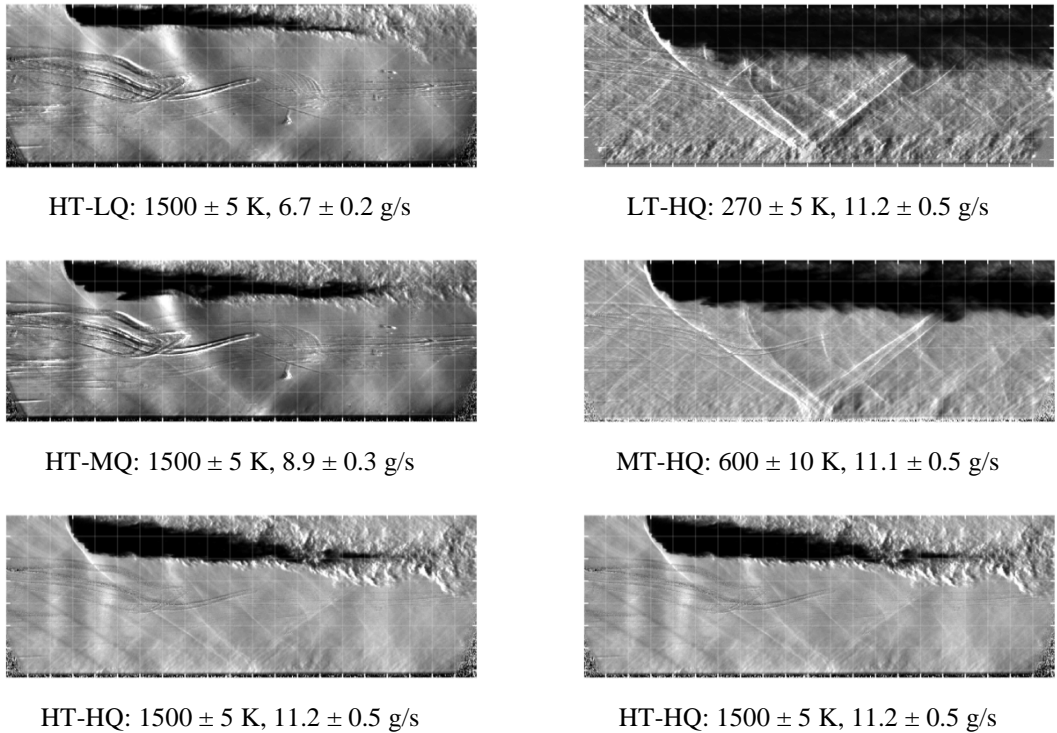

| | |
|---|---|
| HT-LQ: 1500 ± 5 K, 6.7 ± 0.2 g/s | LT-HQ: 270 ± 5 K, 11.2 ± 0.5 g/s |
| HT-MQ: 1500 ± 5 K, 8.9 ± 0.3 g/s | MT-HQ: 600 ± 10 K, 11.1 ± 0.5 g/s |
| HT-HQ: 1500 ± 5 K, 11.2 ± 0.5 g/s | HT-HQ: 1500 ± 5 K, 11.2 ± 0.5 g/s |

**Fig.3** Overview of the tested conditions: instantaneous schlieren images showing the effect of fuel mass flow rate (left) or $T_0$ (right) or on the jet penetration and related shocks. Scale: 5mm

An overview of the test cases is shown on Fig.3 with typical instantaneous schlieren images. A grid is superimposed on each image with a resolution of 5 mm. The kerosene jet is injected on the top of each image and appears in black as shocks are seen as white lines. On the left side of each image, horizontal marks can be seen in the middle of each image. The marks are due to carbon ooze produced by the graphite seal around the optical accesses. The normalization of each instantaneous image by a background without the kerosene jet suppresses well these marks except in the case of HT-LQ and HT-MQ. On the left side of Fig.3, the influence of the jet–to–crossflow momentum flux ratio can be seen from top to bottom, with an increasing kerosene mass flow rate, for the same supersonic flow conditions ($T_0$=1500 K, $\dot{m}_{air*}$ =313 g/s). The influence of q on the spray penetration is clearly visible: the liquid column penetrates vertically further from the injection point as q increases. Indeed, the evaporation becomes more important as $\dot{m}_j$ increases, consequently the mixing is longer to vaporize the liquid jet.
. On the right of Fig.3, images are sorted vertically to see the influence of the supersonic flow temperature, from 270 K to 1500 K, from top to bottom. The influence of the temperature is clearly seen on the horizontal part of the spray plume, which is much longer and optically denser in the lower temperature cases, exceeding the field of view. In the HT-HQ case, the horizontal part of the spray plume is much shorter than in the two other cases HT-MQ and HT-LQ, due to higher vaporization rate. The shocks appears to be more contrasted and visible in the colder cases (LT-HQ and MT-HQ), probably because when the temperature increases, the gaseous density decreases and is thus less perceptible for the Schlieren technique.

Table 1 Main operating parameters of the LAPCAT experiments

| | Supersonic flow | | | | | | | Fuel jet | | | | | |
|---|---|---|---|---|---|---|---|---|---|---|---|---|---|
| Case | $P_0$ (MPa) | $T_0$ (K) | $\dot{m}_{air*}$ (g/s) | $U_{air*}$ (m/s) | $X_{N2}$ | $X_{O2}$ | $X_{Ar}$ | $X_{H2O}$ | $P_{inj}$ (MPa) | $\dot{m}_j$ (g/s) | $u_j$ (m/s) | q | $We_{air*}$ | Re |
| LT-HQ | 0.4 | 270 | 762 | 482 | 0.781 | 0.209 | 0.009 | 0.000 | 2.15 | 11.2 | 55.6 | 8.6 | 6533 | 15763 |
| MT-HQ | 0.4 | 608 | 529 | 752 | 0.763 | 0.182 | 0.009 | 0.045 | 2.13 | 11.1 | 55.1 | 7.9 | 6926 | 15621 |
| HT-HQ | 0.4 | 1501 | 314 | 1233 | 0.584 | 0.210 | 0.007 | 0.198 | 2.15 | 11.2 | 55.6 | 8.6 | 6551 | 15763 |
| HT-MQ | 0.4 | 1505 | 313 | 1233 | 0.584 | 0.210 | 0.007 | 0.199 | 1.38 | 8.87 | 44.0 | 5.4 | 6551 | 12474 |
| HT-LQ | 0.4 | 1500 | 314 | 1232 | 0.584 | 0.210 | 0.007 | 0.199 | 0.78 | 6.70 | 33.2 | 3.1 | 6540 | 9412 |

## 2.2 High-speed shadowgraphy

A high-speed shadowgraphy setup is first used to visualize the droplets location in the spray in order to explore the whole atomization mechanism. The shadowgraphy system is complementary to the schlieren setup, described in the next section, because it allows a better identification of the area where the liquid is present, as it less sensible to density gradients. The spray was illuminated in a backlight configuration by a high-magnitude light source Prolight 575 W, providing a white incoherent and continuous light. A Fresnel lens and a transparent sheet were placed in front of the light to ensure a homogeneous background on images. The dark level of the camera was subtracted for each run and images were normalized to suppress background non uniformity.

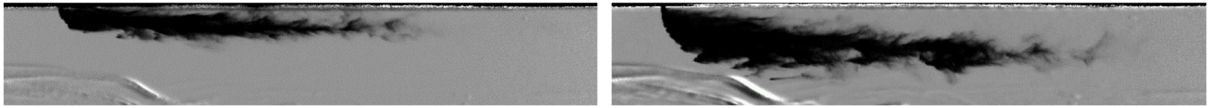

**Fig.4** Instantaneous shadowgraphs obtained for two jet mass flows: 6.7 g/s, HT-LQ (left) and 11.2 g/s, HT-HQ (right)

Images of the spray were recorded with a Phantom v711 high-speed camera from Vision Research. This camera is composed of a 12-bit CMOS sensor of 1280 x 800 square pixels and of 20 μm side length. The resolution of the sensor is directly related to the frequency rate. The maximum frame rate is 7500 Hz when the image resolution is set at its maximum of 1280 x 800 pix². An objective of 105 mm focal length was mounted on the camera and the aperture was fixed to f/5.6 during the whole tests. Typical shadowgraphs presented on Fig.4 are cropped to 896 x 152 pix², providing a field of view of 101.9 mm x 17.4 mm, with a spatial resolution of 8.8 pix/mm. On Fig. 4, the influence of the flow rate on the jet penetration can be observed on these two instantaneous images: in the HT-LQ case, the penetration is shorter than in HT-HQ case, with some liquid probably flowing on the upper wall. On the bottom left corner of each image, some artefacts can be seen due to several marks on the window between the camera and the spray than cannot be totally removed by the normalization.

The camera exposure was set to 1 μs to limit the blurring effect due to the high velocity of the droplets. Nevertheless, a pixel shift on droplet images was estimated to ≈1 pixel for a droplet velocity of 100 m/s. As the goal of the shadowgraphy is to measure the penetration height of the spray, this pixel blur is considered as negligible. Time-averaged images are calculated over 400 images during the stationary phase of the flow to measure the penetration of the liquid jet.

## 2.3 High-speed schlieren

The schlieren imaging method, as described by Desse and Deron, 2009, involves intercepting, with a knife, some of the light rays which are deflected in the test section. Schlieren is used as an imaging technique to visualize the density gradient in a flow resulting from deviations through the optical path of the light rays crossing the supersonic flow. In this experiment, schlieren allows to visualize the shock structures as white lines, the fuel vapours and turbulent mixing area in grey level gradients and the spray in black. The schlieren setup consists in forming a small punctual light source from a condenser (1 mm in diameter) and cut by a knife. The white and continuous light source is placed at the 800 mm focal distance of a Clairaut lens to produce a parallel light of ≈120 mm in diameter. A part of the light rays passing through the test section are intercepted by a knife edge while the other part is focused on the camera. The high-speed schlieren imaging setup is set with a lower spatial resolution but a higher temporal resolution, compared to shadowgraphy. The optical setup has a spatial resolution of 6.1 pix/mm. A first set of images is recorded at 5 or 10 kHz with an image size of 800 x 256 pix² to visualize the whole penetration of the liquid jet. Images presented on Fig.3 and Fig.5 are normalized and cropped at 640 x 224 pix², to obtain a resulting field of view of 105 mm x 37 mm. A second set of images, recorded at a higher frame rate, up to 210 kHz for 128 x 128 pix², allow to follow the convectional flow structures and the temporal dynamics of the shocks oscillations but also to measure the jet velocity, at the same spatial resolution of 6.1 pix/mm. On the typical schlieren images presented on Fig.5, the main physical processes of the liquid injection in a supersonic crossflow can be observed: the shocks and their reflexions on the walls can be seen as white lines and the liquid jet as dark pixel values. All schlieren images on Fig.3 clearly show the bow shock, located at approximately a 0.3 mm upstream from the base of the liquid jet. The shock induced boundary layer separation is less intense than the bow shock and can only be seen on the colder test cases (LTHQ and MTHQ on Fig.5), due to a higher camera exposure time (1 µs instead of 0.5 µs). Moreover, the position of the leading edge shock can be seen as a diagonal line upstream the bow shock.

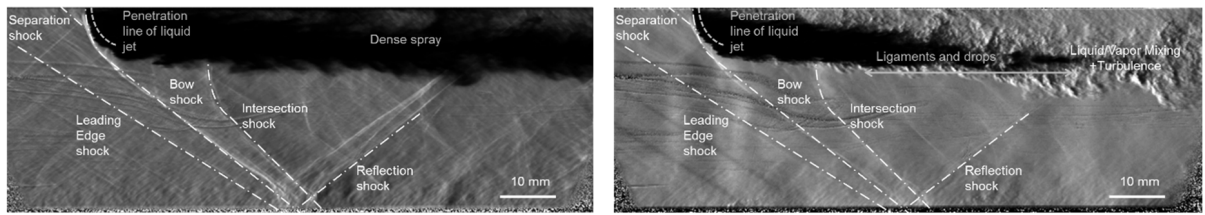

**Fig.5** Instantaneous schlieren images. Cases MT-HQ (left) and HT-HQ (right)

When a supersonic meets an obstacle, a shock is generated upstream from this obstacle. In the case of a jet injection, a so-called bow shock appears a few millimetres upstream, following the base of the liquid column. This shock turns to an oblique shock once the liquid jet has penetrated deeper in the chamber. The bow shock originating at the base of the injection point appears to follow the liquid column in the first millimetres and the penetration line, illustrated by a white arc in dashed line. On the left side of the liquid column, namely the windward region, surface instabilities can be seen as small waves travelling along the interface between the fuel jet and the crossflow (i.e. left side of the liquid column). To the right of the injection point, namely the leeward region, the spray is too dense optically to see the interface and the breakup of the liquid column. Following the bended liquid column, once past a breaking point, which cannot be seen in this image, ligaments and drops are created in the jet plume. At the right end of the image, an area with strong fluctuations of grey levels indicates density gradients due to turbulence and kerosene vapour and liquid mixing with the supersonic flow.

## 2.4 High spatial resolution imaging

A high spatial resolution imaging system has been set in order to resolve the smallest sizes of the atomization process. In this shadowgraphy setup, the spray is lit with a backlight configuration by a laser diode CAVILUX Smart 400W which emits a red incoherent light pulse at 640 ± 10 nm. This source is very compact and stable in space and time, providing light through an optical fiber. In our application where droplets are very fast, the pulse duration was set to 40 or 60 ns to freeze their motion on the images and reduce the blurring effect. A Fourier lens of 40 mm in focal length is placed at the fiber exit, as well as a diffuser to create a homogeneous illumination area. As usual, images are normalized to suppress background inhomogeneity such as the area close to the injector exit which is in the shadow of the chamber wall.

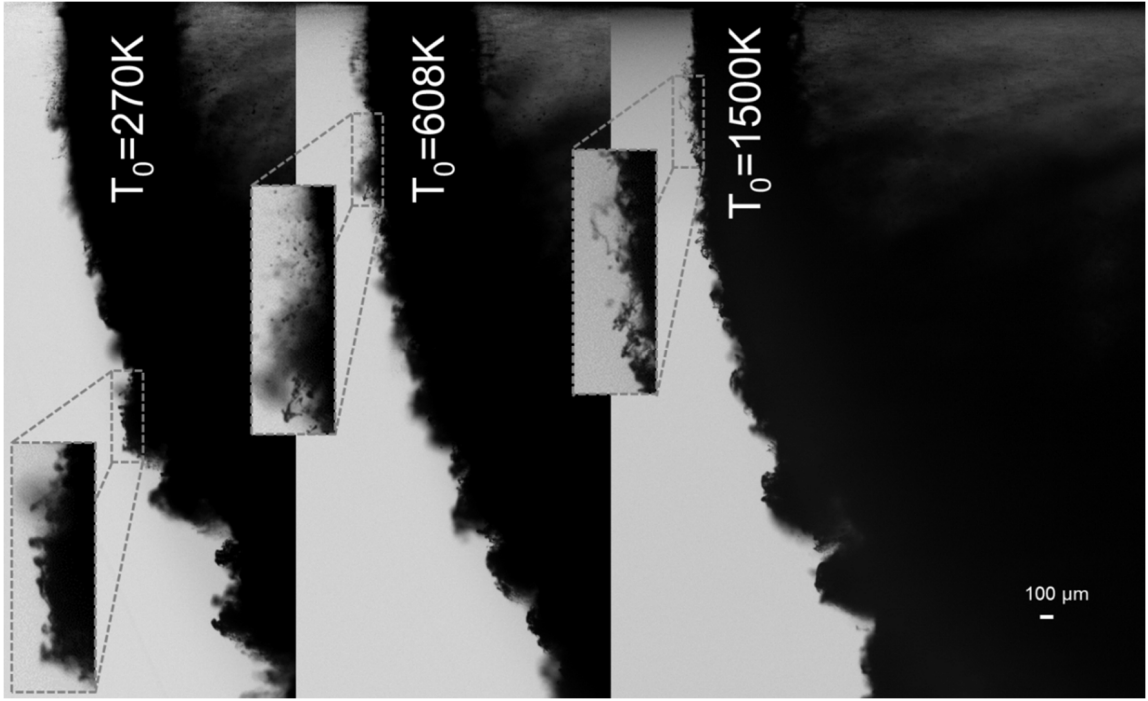

**Fig. 6** Instantaneous high-resolution images for three given temperature, windward side of the jet

Images of the spray were recorded with an AcA4112-30um camera from Basler. This camera is composed of a 12-bit CMOS sensor of 4112 × 3008 square pixels and of 3.45 μm side. The camera is equipped with a long working distance microscope K2 Distamax (Infinity) of lateral magnification $g = 4.7$ is used to obtain high-resolution images. The field of view is 5.6 mm x 4.1 mm and the resolution is 733 pixel/mm. The repetition rate was fixed by the camera and set at 25 Hz. Due to the spatial resolution, one pixel displacement during the light pulse exposure duration of the light pulse (60 ns) induces a velocity of 23 m/s. In these conditions, it is not possible to measure droplet diameter from images. The optical system is mounted on a 3D displacement system to record images at different positions in the spray. Typical high-resolution images are presented on Fig. 6 showing the liquid jet interface for the three studied temperatures.

3 Results

3.1 Atomization mechanism

The atomization of a liquid jet submitted to an air stream has a main influence on the fuel/air mixing and thus ignition and combustion. Beloki et al., 2009, propose a comprehensive schematic diagram of a liquid jet atomized by a supersonic cross flow. The atomization in such a supersonic flow is led by primary and secondary breakups physical processes, and the relative importance between Kelvin-Helmholtz instability and Rayleigh-Taylor instabilities is still under discussion (Ren et al. 2019). Indeed, as the liquid fuel is injected in the gaseous flow, surface wave instabilities (Kelvin-Helmholtz) develop at the liquid-gas interface, as seen on the left side of the liquid column on Fig. 6. These waves are pushed and amplified by the gaseous flow, resulting in the formation of very thin and short ligaments, leading to droplets by Rayleigh-Taylor instabilities, illustrated in the magnified region of interest on the left side picture of Fig.6. The breakup of the liquid jet into ligaments and droplets is defined as the primary breakup. These ligaments are rapidly peeled off the jet, stretched and sheared by the difference of velocity between the liquid jet and the outer gas stream and finally broken into smaller droplets: this process is called secondary breakup. The mass shedding resulting from the secondary atomization process cannot be seen along the liquid column on Fig.6, as the liquid elements are quickly blown away by the supersonic flow. Nevertheless, they can be seen downstream as detached elements on Fig.4. This description of the atomization process almost corresponds to the classical atomization described in the case of a liquid jet injection in a subsonic cross-flow. The breakup of the liquid jet is in a shear breakup regime, according to the We, q correlation maps listed in Broumand and Birouk, 2016.

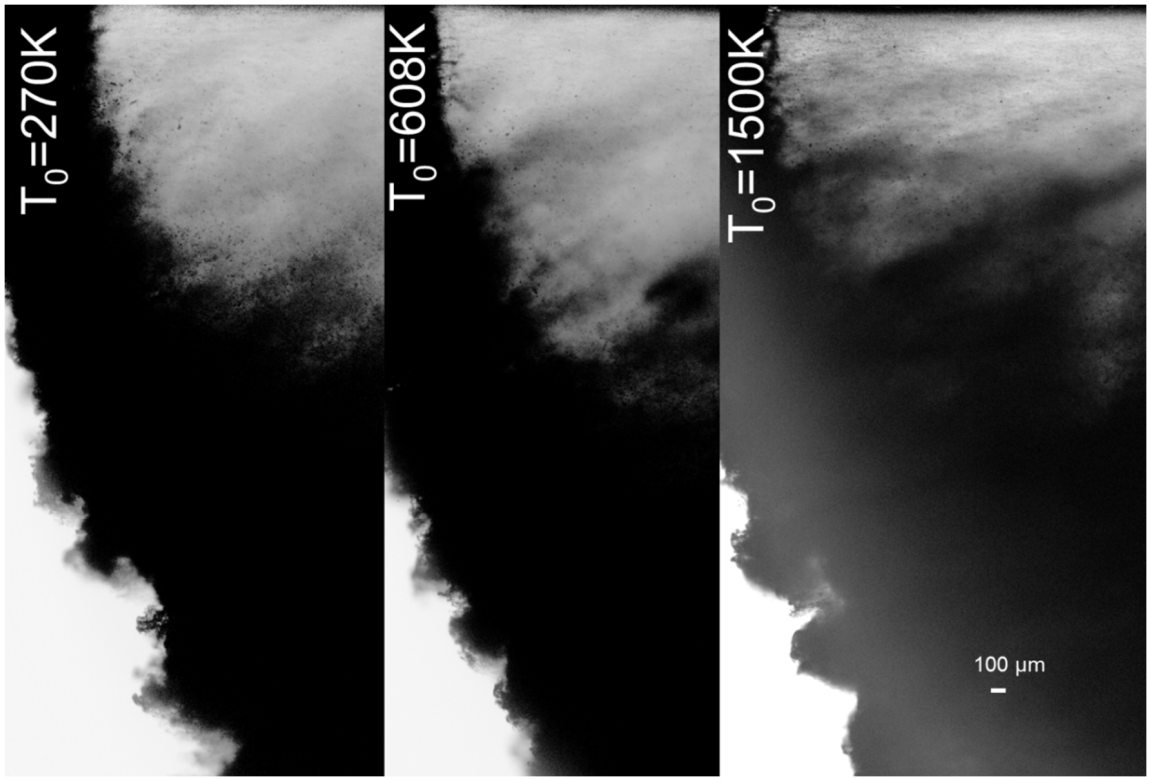

**Fig. 7** Instantaneous high-resolution images for three given temperature, leeward side of the jet

In the case of the supersonic flow, the atomization process is more complex due to the presence of the bow shock and shock induced boundary layer separation (Beloki et al., 2009). As the round liquid fuel jet penetrates into the supersonic crossflow, a bow shock forms ahead of the jet. Meanwhile, the inverse pressure gradient in the streamwise direction induces a recirculation zone upstream of the inlet of the jet and causes the separation of the boundary layer. Consequently, there is a production of small droplets at the nozzle exit that can be seen on the magnified pictures of Fig.6, probably resulting from this shock interaction, hiding the surface waves at the liquid column base. The order of magnitude of the droplet size produced in this area, linked to the width of the ligaments is estimated at about 5 to 10 µm, which is a little smaller, compared to droplet size measurements studies at Ma≈2 of Lin et al. 2004, with water or Wei et al. 2020, with kerosene jets.

The characteristic length scale of the Kelvin-Helmholtz instability, evidenced by the surface waves along the liquid column, can be estimated in the middle of the instantaneous images of Fig. 6. The wavelength has been measured on the windward side of the liquid jet on several successive waves for the three temperature conditions considered in this study, on several pictures such as the one presented on Fig.6. The estimated wavelength λ is about 0.5 mm, 0.4 mm and 0.55 mm, for LT (270 K), MT (608 K) and HT (1500 K) respectively. Thus λ seems to increase when $We_{air*}$ decreases. This tendency is consistent with the scaling law of Sallam et al. 2004, $\lambda/d_j=C_1 We_{air}^{-0.45}$, where $C_1$ is a constant, $We_{air}>4$ and $3<q<8000$.

The droplets produced are very small in sizes, several orders of magnitude smaller than the diameter of injection. These droplets are rapidly transported by the supersonic flow from the location they appear to the leeward region of the spray. Indeed on Fig. 7, instantaneous high-resolution images enhancing the leeward region of the spray show streaks of droplets directed almost perpendicularly to the liquid column surface. The trajectories of these streaks are driven by the jet plume interaction with the supersonic cross flow, inducing a complex vortical structure (Viti et al., 2009). Indeed, a counter-rotating vortex pair (also referred to as counter rotating trailing vortices, Pelletier et al., 2021) develops along the average trajectory of the liquid jet plume, starting at the leading edges of the jet shear-layers. As the vitiated airstream approaches the jet, the boundary layer separates, leading to the formation of a horseshoe vortex. The formation of this vortex is accompanied by a separation shock associated to a pressure increase, which is followed by a pressure decrease in the vicinity of the vortex. The droplets are thus carried from the windward side of the leeward side of the liquid column and trapped by these vortices developing on the side of the shear-layer jet.

3.1 Penetration depth

The penetration of the liquid jet into the supersonic flow may be an important feature when designing a liquid-fuelled scramjet combustor. A deep penetration may be desired to increase the mixing and thus lower the ignition distance. One may also wish to avoid spray impingement onto the combustor walls. Since 1970, many experimental studies can be found in the literature with various injection configurations for subsonic (Broumand and Birouk, 2016), but also supersonic flows (Ren et al., 2019). In such a situation, the momentum flux ratio q is the main parameter that drives the penetration length. The penetration height Y is defined as the liquid maximum penetration perpendicularly to the crossflow, from the injection point. The penetration measurements presented here are obtained in single jet injected in a supersonic flow with $3.1<q<8.6$ and $270\ K<T_0<1500\ K$; and compared to other studies with a similar injection setup.

Time-averaged schlieren images are presented on Fig. 8, on the left side images are sorted by increasing q from top to bottom. The contour of the liquid plume obtained from the binarization of shadow images are superimposed as a white line to schlieren images in the background. The contour obtained with time-averaged shadowgraphs is rescaled to correspond to the schlieren image resolution. The liquid jet plume obtained with shadowgraphs ensures an accurate binarization as the resolution is higher and images are well contrasted compared to the background as the limit between liquid and gas is clearer. Nevertheless, the contours of the image plume matches well with schlieren images which confirms that both imaging techniques can be used to measure the penetration. As already seen on Fig.3, the liquid column penetrates further from the injection point as q increases. Indeed, as $\dot{m}_j$ increases, the evaporation becomes more important, consequently, the mixing is longer to vaporize the liquid jet.

On the right side of Fig. 8, time-averaged schlieren images show the influence of the supersonic crossflow temperature $T_0$, with temperature increasing from top to bottom. The jet plume is very dark in the leeward direction in the LT-HQ and MT-HQ cases. The effect of vaporization is evidenced comparing these two cases with HT-HQ, as the plume becomes thinner but doesn't seem to penetrate deeper (in the y direction). The bow shock position seems to be more visible for the low temperature cases, which could be due to a better sensibility of the schlieren setup as ρ decreases with increasing $T_0$ but also because the alignment is set at ambient temperature.

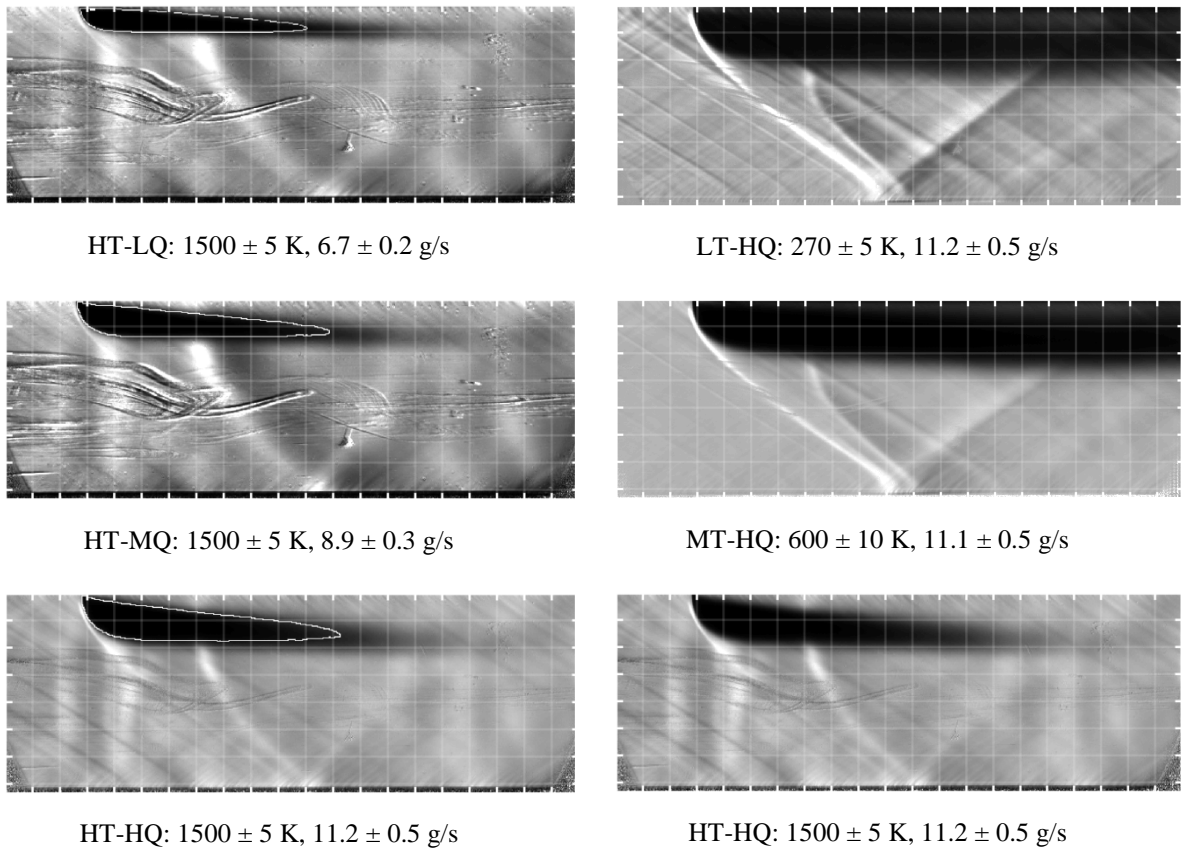

**Fig.8** Time-averaged schlieren images showing the effect of fuel mass flow rate (left) or $T_0$ (right) on the jet penetration and related shocks. Scale: 5mm

An image processing algorithm has been developed to measure the spray penetration height $y/d_j$ from images of Fig. 8. It is based on the binarization of the time-averaged image for each operating condition of Table 1. The algorithm is the same for shadowgraph and schlieren images. A set of instantaneous normalized images recorded during the stationary phase of the flow is averaged. The resulting time-averaged image is then binarized to separate the denser part of the liquid phase of the gaseous and dispersed phase. A relative threshold is then defined from the grey level histogram thanks to the "Minimum method" originally proposed by Prewitt and Mendelsohn, 1966 and implemented in Image J (Fiji distribution, Schindelin et al., 2012). Once the binarized image is obtained, the maximum position of the black pixels, perpendicularly from the upper wall is extracted to calculate the penetration height, relatively to the injector position.

On Fig. 9, dimensionless penetration heights $y/d_j$ versus the longitudinal distance from the injection point $x/d_j$ are presented for the different test cases of Table 1. Axial and vertical distances are normalized by the injector orifice diameter $d_j$ reported in Table 2. The effect of temperature on the penetration is shown on Fig. 9.a, with grey-scale level of curves getting darker as $T_0$ increases. The lower temperature conditions LT-HQ and MT-HQ are compared to the well-established correlations of a research group that has conducted many experimental studies on the spray structures of liquid jets in supersonic cross flows from the early 2000, providing the penetration correlations in Table 2, obtained with either shadowgraphy (Lin et al., 2002) or PDPA (Lin et al., 2004). The correlation proposed by Lin et al. 2002 was obtained for different liquids (water, alcohol or a mix of water and alcohol) injected in a Mach 1.94 supersonic flow at ambient temperature (293 K), which is closed to the case LT-HQ. The penetration measured in the case LT-HQ (light grey solid curve) and the Lin et al., 2002 correlation (dot/dash line) shown on Fig. 9.a are in a relatively good agreement. In Lin et al., 2004, pure water is injected through a single injector in a supersonic flow (Mach 1.94), at a total temperature of 477 K and 533 K. These measurements were performed in the range $25<x/d_j<200$ and $2<q<15$, which is close to the presented MT-HQ case (medium grey curve), except for the injected liquid (water instead of kerosene). On Fig. 9.a, the penetration height predicted by the correlation obtained with PDPA (dash line) is higher than the MT-HQ case (medium grey curve) and that predicted by correlation derived from their previous work from shadowgraph images (Lin et al. 2002). According to Lin et al., 2004, the correlation obtained with PDPA overestimates the penetrations obtained with shadowgraph by 40%, because shadowgraph images show primarily droplet number density and may not show the upper portion of the spray plume, which contains large droplets at low number density. The penetration trajectories obtained in the present study are about 20% lower than their work, at x/d=10, which can not be explained only by the different measurement methods bringing to this correlation. The choice of the threshold to obtain the penetration has a non negligible influence in imaging methods and in our case, doubling the threshold value induces an increase of the penetration of 10%. Another explanation could be the difference in the fluid injected. Indeed, the dynamic viscosity of kerosene is 1.9 times higher than the water viscosity at ambient temperature. Amongst the authors who studied the influence of the viscosity on the penetration of various fluids injected in a subsonic crossflow in different conditions, Stenzler et al., 2006 reported a dependency of the penetration with the viscosity ratio to water such as $(\mu_j/\mu_{water})^a$ where $0.027<a<0.108$. It means that in our case, the penetration length could be reduced of about 5% due to the fuel viscosity.

No comparison with literature can be done with HT-HQ case as no work reported penetration measurements in such a high temperature supersonic flow with kerosene jet injected. Comparing the three present test cases of Fig.9.a, the effect of the temperature increase shows a decrease of the penetration. This is in agreement with Lakhamraju and Jeng, 2005 who reported that at constant q and $We_{air*}$, jet penetration decreases with increasing crossflow temperature. Indeed, when q and $We_{air*}$ are constant, like in the cases of LT-HQ and MT-HQ, the gas inertial forces $\rho_{air*} u_{air*}^2$ and the liquid jet inertial force $\rho_j u_j^2$ must be constant, thus an increase in temperature induces a decrease of $\rho_{air*}$ and hence an increase of $u_{air*}$. Since the exponent of $\rho_{air*}$ is smaller than that of $u_{air*}$ in the correlation of the rate of mass shedding, a decrease in $\rho_{air*}$ and an increase in $u_{air*}$ will lead to an increase in mass shedding, creating smaller droplets and therefore leading to a lower jet penetration (Broumand and Birouk, 2016). Finding a correlation to represent the effect of the temperature on the penetration of the liquid jet was not possible due to the limited number of test runs and the physical processes involved in these test cases. Indeed, HT-HQ case is different from LT-HQ and MT-HQ as the temperature is higher than the kerosene evaporation temperature. The vaporization effect will also lead to creation of smaller droplets and penetration.

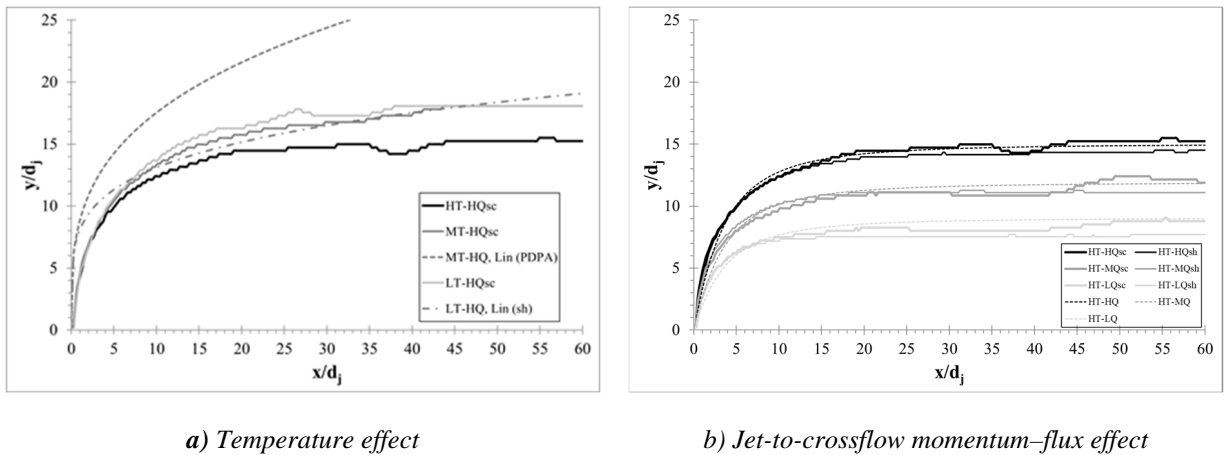

*a) Temperature effect*  *b) Jet-to-crossflow momentum–flux effect*

**Fig.9** penetration height as a function of axial distance from the injection point, sc and sh stand for schlieren and shadow images. Dot/dash lines refers to correlation in Table 2

The influence of the momentum flux ratio q can be seen on Fig.9.b by comparing HT-LQ, HT-MQ and HT-HQ, with increasing q, for a given total temperature of about 1500 K. Solid lines refer to the present experiment, with the solid line colour getting darker as q increases. Penetration measurements obtained with schlieren images are indicated as 'sc' with bold lines and with shadowgraphy are indicated as 'sh' with thinner lines. Measurements performed with shadowgraphy and schlieren images are in a good agreement, for a given operating condition (comparing bold and light lines of the same colour together). This ensures the robustness of the algorithm, particularly with the automatic binary threshold choice. This also gives an indication of the satisfactory reproducibility of the test conditions from a run to another. Additionally, a correlation for the penetration of the kerosene jet injected in the high temperature supersonic flow is indicated as dot lines on Fig. 9.b and reported in Table 2. As reported by many authors, $q^{\alpha}$, with $0.4<\alpha<0.5$ (Lin et al., 2002, Ghenai et al., 2009) is the main parameter that drives the penetration in a pure liquid jet injected in a supersonic flow, thus $\alpha=0.5$ is chosen for this study. Nonetheless, one can note that the curves tend to a limited height $[y/d_j]_{max}$. Consequently, the following correlation is proposed: $y/d_j=[y/d_j]_{max}\tanh[\ln(1+\beta x/d_j)]$, with $\beta$ and $[y/d_j]_{max}$ two functions determined by the main parameters of the flow (q, $We_{air*}$ and the viscosities of liquid and fuel).

**Table 2** Correlations of the liquid jet penetration in a supersonic crossflow

| Correlations | Ma | T0 (K) | Liquid | $d_j$ (mm) | q | $x/d_j$ | Reference |
|---|---|---|---|---|---|---|---|
| $Y/d_j=4.73q^{0.3}(x/d_j)^{0.3}$ | 1.94 | 477-533 | $H_2O$ | 0.5;1.0 | 2 - 15 | 25-200 | Lin et al., 2004 (PDPA) |
| $Y/d_j=3.94q^{0.47}(x/d_j)^{0.21}$ | 1.94 | 293 | $H_2O$, alcohol | 0.5-2.0 | 1 -18.5 | 0-85 | Lin et al., 2002 (Shadowgraph) |
| $Y/dj=5.13q^{0.5} \tanh[\ln(1+0.25x/d_j)]$ | 1.95 | 1500 | Kerosene | 0.567 | 3.1-8.6 | 0-60 | Present study |

3.2 Jet velocity

A PIV software was used to measure the velocity of liquid elements along the liquid jet from shadowgraphy images, as in the work of Beloki et al., 2009. Velocity fields can be obtained if the image frequency is high enough to capture the displacement of the liquid jet on few pixels. In our work, the PIV algorithm was applied on high-speed schlieren images recorded at 210526 Hz to track the structure motion, mainly at the interface of the liquid jet with the gaseous flow. Results presented in this paper are calculated from 25000 image couples.

FOLKI-SPIV was developed at Onera by Champagnat et al. (2011). It is an extension of FOLKI (Le Besnerais and Champagnat, 2005), which is an optical flow software, based on the Lucas-Kanade paradigm (Lucas and Kanade, 1981). FOLKI-SPIV is found comparable to that of state-of-the-art FFT-based commercial PIV software, while being faster. Indeed, the computation of dense PIV vector fields uses the graphics processing units (GPUs) resources of the computer. This method is based on an iterative gradient-based cross-correlation optimization method, which provides an accurate and efficient alternative to multi-pass processing with FFT-based cross-correlation. Density is meant here from the sampling point of view (one vector per pixel is obtained), since the presented algorithm, FOLKI, naturally performs fast correlation optimization over interrogation windows with maximal overlap. On the instantaneous image couple presented on Fig. 10, representing a field of-view of 12.5 mm x 10.5 mm, liquid structure displacements can be followed along the liquid column. The algorithm confidence is evaluated with the normalized cross-correlation score as in a standard PIV software. At the bottom left of Fig. 10, the score is ranging from 0 (no correlation at all, in black) to 1 (perfect correlation, in white). The score was found to be high enough (typically S>0.5) in the area of the liquid/gas interface along the liquid column and the bow shock. Everywhere else, the score is too low to provide a reliable velocity value.

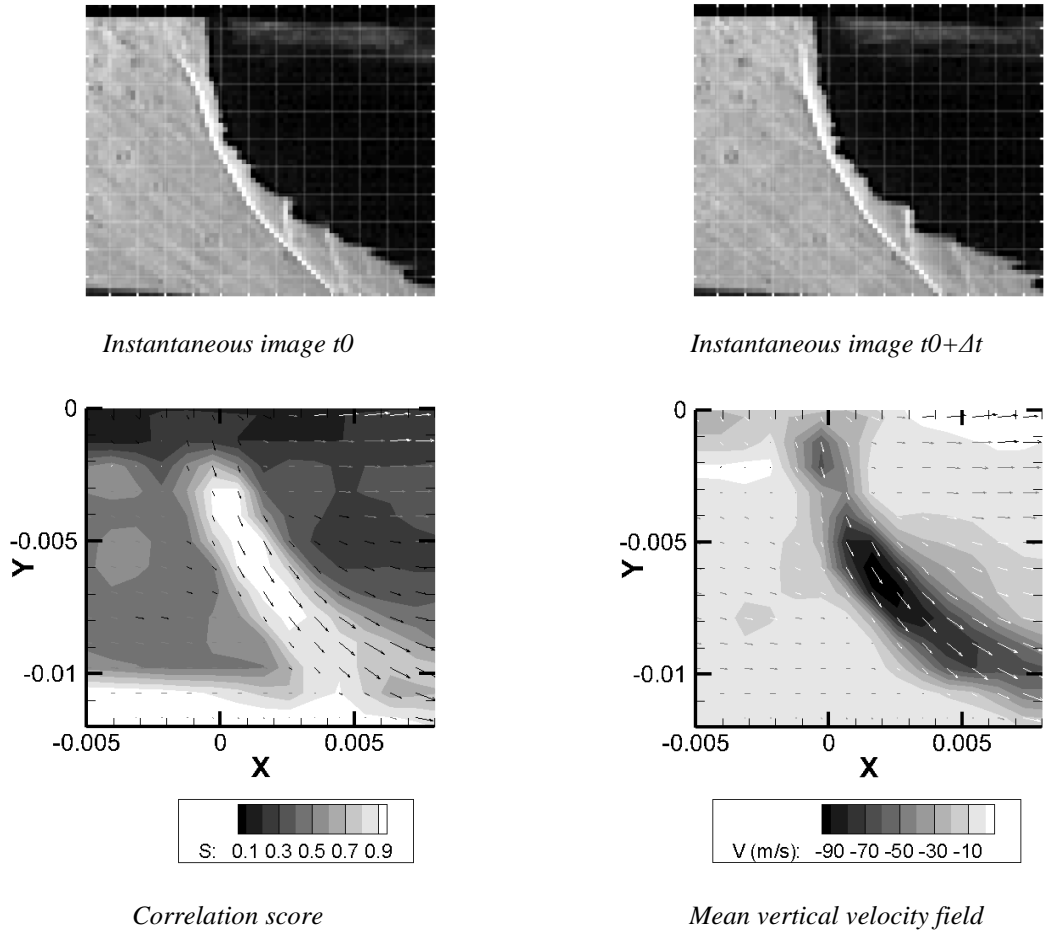

**Fig.10** Instantaneous schlieren image couple (top, scale: 1 mm) and
time-averaged correlation score and velocity fields in m/s (bottom), MT-HQ

FOLKI-SPIV provides one velocity vector per pixel, but the effective spatial resolution remains linked to the interrogation window size, as for conventional PIV methods. The optimal interrogation window size in term of cross-correlation score was found to be $9 \times 9$ pixel², which is large regarding the raw image size ($128 \times 128$ pixel²). As images were not homogeneously filled with liquid structures, a large interrogation window size enabled to obtain more reliable results in areas suffering from a lack of liquid structures. Increasing the interrogation window size provides larger spatial scales but it also decreases the signal to noise ratio. The camera exposure time is set to 0.5 µs, which is about 10% of the time interval between images at 210 kHz ($\Delta t$=4.75 µs). Thus an uncertainty of the same amount has to be considered on the measured velocity.

At the bottom right of Fig. 10, the grey scale for the vertical velocity indicates maximum negative velocities in black, directed from top to bottom. Positive vertical velocities (in white) can be found upstream from the liquid jet, close to the injection wall. In this area, the correlation score is weak so the velocity field is questionable. In the MT-HQ operating condition, the vertical velocity at injection estimated from the fuel mass flow rate through the nozzle diameter is 55.1 m/s (see Table 1). At the injection point (x=0, y=0), the score is too weak to provide a reliable vertical velocity. At the closest position of the injection point where the score is higher than 0.5, that is x=-0.5 mm, y=-2 mm, the vertical velocity computed by FOLI-SPIV is -55 m/s, which is in agreement with the velocity estimated at injection.

The liquid column is located about half a millimetre downstream, behind the bow shock. The separation shock can also be seen on the left of the instantaneous schlieren image of Fig. 10, close to the injection point, at the left of the liquid jet. Moreover, small shocks can also be seen between the main bow shock and the liquid waves along the liquid column. These "microshocks", also evidenced by Sathiyamoorthy et al. 2020 on tandem liquid jets injection configurations in supersonic crossflow, are following every wave travelling along the liquid column. Due to a lack of spatial resolution, it is hard to say if they emerge from the interaction between the separation shock and the bow shock or a little further downstream from the liquid wave itself.

### 3.2.1 Effect of fuel mass flow rate on velocity field

The set of velocity fields presented on the left of Fig. 11 shows the effect of the fuel mass flow rate on the mean vertical (right) velocity fields, with increasing q from top to bottom. The measured vertical velocity near the injection point is in agreement with estimated velocity at the injection point (x=0, y=0). Indeed, for HT-LQ, HT-MQ and HT-HQ, the estimated bulk velocity through the injector is respectively 33 m/s, 44 m/s and 55 m/s. The measured vertical velocity at the closest position from the injection point where the score is high enough to be reliable (S≈0.6) is -34 m/s, -50 m/s, -63 m/s respectively for HT-LQ, HT-MQ and HT-HQ at x=-0.5 mm, y=-2 mm. This values are in agreement with the estimated velocities, considering the uncertainty of the velocity measurements, thus the velocities measured with the PIV software are close to the one of the liquid jet. As the vertical velocity near the injection point is increasing, the liquid jet penetrates further from the injection plane. The vertical kinetic energy is rapidly transferred to horizontal kinetic energy, as the velocity vectors are almost horizontal before exiting the field of view.

### 3.2.3 Effect of temperature on velocity field

On the right of Fig.11, mean vertical velocity fields are presented for the three temperature conditions, with increasing $T_0$ from top to bottom. Even if the gaseous velocity $u_{air*}$ is increasing when the temperature increases, it doesn't have a direct impact on the measured liquid jet velocities. Indeed, when comparing LT-HQ and MT-HQ, q and $We_{air*}$ are kept constant, thus an increase of $u_{air*}$ is compensated with a decrease of $\rho_{air*}$ due to the flow temperature increase. The ratio of inertial forces of the liquid and the gas is then constant and the velocity field remains similar. For HT-HQ case, the gaseous velocity $u_{air*}$ is higher by increasing the speed of sound, inducing a lower q, and giving a higher inertia to the crossflow compared to the jet. But as $We_{air*}$ is smaller than the other cases, due to the temperature increase, the crossflow inertial force is relatively lower, inducing smaller velocities before the liquid jet exits the field-of-view.

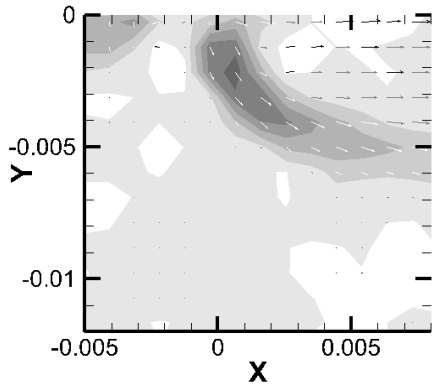
*Mean vertical velocity field, HT-LQ*

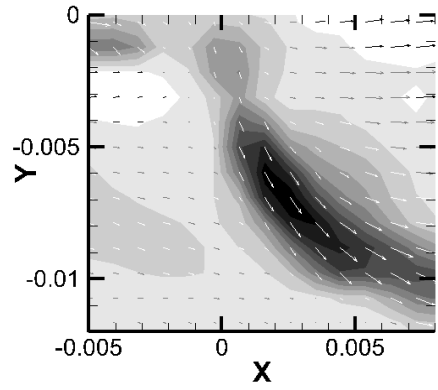
*Mean vertical velocity field, LT-HQ*

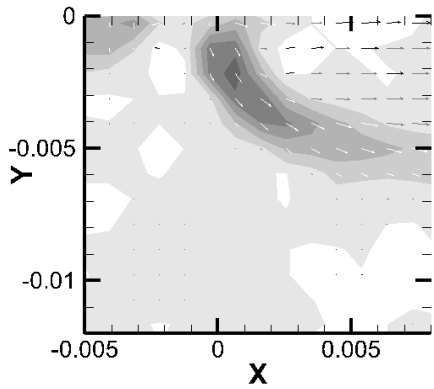
*Mean vertical velocity field, HT-MQ*

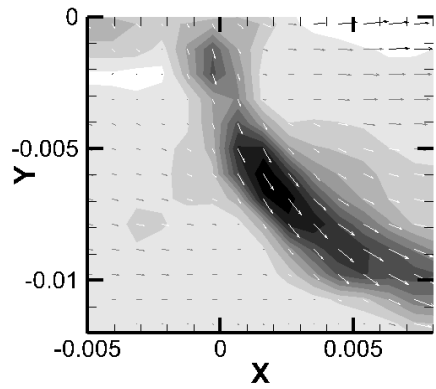
*Mean vertical velocity field, MT-HQ*

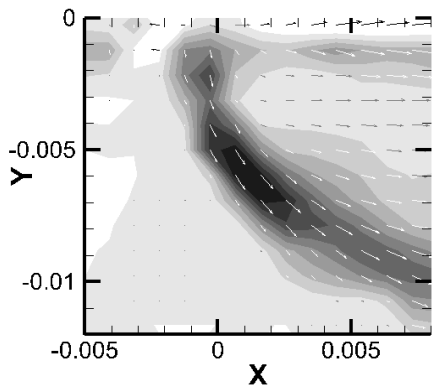
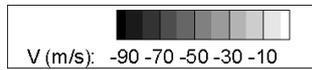
*Mean vertical velocity field, HT-HQ*

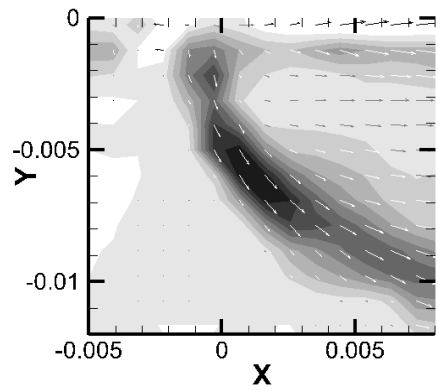
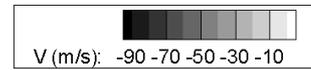
*Mean vertical velocity field, HT-HQ*

**Fig.11** Mean vertical velocity fields of the liquid structures in m/s, influence of q (left) and $T_0$ (right)

4 Conclusions

The deep penetration of atomized liquid fuel into the supersonic airflow is a key physical process to maintain the combustion. There is a lack of experimental studies in conditions close to engines operation (i.e. with real fuel and hot air flow). Atomization and penetration of a kerosene jet injected in a Mach 2 crossflow were studied experimentally in the LAPCAT-II Dual Mode Ramjet Combustor at Onera, using high spatial and temporal resolution imaging techniques. This experimental study presents measurements in high temperature conditions, with kerosene as liquid fuel. The fuel spray is injected through a single orifice, perpendicularly to a Mach 2 supersonic air flow, at temperatures from ambient to 1500 K and jet-to-crossflow momentum flux ratio from 3.1 to 8.6. Five different test cases have been investigated to characterize independently the effect of stagnation temperature $T_0$ or q on the atomization and mixing process. A high spatial resolution imaging system was used to capture the features of the liquid jet atomization process in the supersonic crossflow. A production of small droplets at the nozzle exit, resulting from the interaction between the bow shock and separation shock has been evidenced on the magnified pictures. The order of magnitude of the droplet size produced in this area, linked to the width of the ligaments is estimated at about 5 to 10 µm, which is a little smaller, compared to droplet size measurements studies in similar conditions. The characteristic length scale of the surface waves along the liquid column has been estimated on the windward side of the liquid to half a millimetre. In the leeward region of the spray, high-resolution images also show streaks of droplets oriented almost perpendicularly to the liquid column surface. The trajectories of these streaks are driven by the jet plume interacting with the bow- and lambda-shocks developing overs the injectors. Such magnified visualizations bring a new insight of the atomization of the liquid jet in the supersonic flow, as suggested by Lin et al. 2020.

An image processing algorithm was developed to measure the penetration height of the spray and applied in the same way to both shadowgraphs and schlieren images. Penetration measured with shadowgraphy and schlieren images are in a good agreement, for given operating conditions. This ensures the robustness of the algorithm but also gives an indication of the reproducibility of the experimental conditions. Penetrations of the kerosene jet are measured for five test cases and compared to other studies of the literature, in the lower temperature cases. The effect of the temperature increase shows a decrease of the penetration due to an increase in mass shedding, creating smaller droplets. Schlieren images are used to detail the complex shock wave structures in such a supersonic flow. Small shocks have also been evidenced between the main bow shock and the liquid waves along the liquid column. These small shocks seem to follow every wave travelling along the liquid column. High-speed schlieren performed at 210 kHz is also performed to capture the temporal dynamics of the shocks and measure the liquid jet velocity.

A PIV software developed at Onera was used to measure the velocity of structures motion along the liquid jet and has been applied to shadowgraphy and schlieren images. Horizontal and vertical velocity fields along the liquid column in various operating conditions are provided, which constitute important data for numerical simulations. The effect of the mass flow rate on the velocity field is investigated and shows the importance of the jet-to-crossflow momentum ratio in the vertical kinetic energy transfer to horizontal kinetic energy. The effect of the temperature on the velocity field is more complex and depends not only on the jet-to-crossflow momentum ratio but also on the relative gaseous Weber number of the operating conditions.

Perspectives of this study are to quantify more precisely the surface wave evolution and characteristic scales along the liquid column, thanks to the high-resolution images. This could explain the relative contribution of the Kelvin-Helmholtz and Rayleigh-Taylor instabilities. The birthplace of the microshocks following the surface wave could be evidenced with complementary experiments with high spatial resolution schlieren imaging. Comparisons between these experimental results with numerical simulations performed at Onera are going on to evaluate different strategies to recover the jet penetration and the velocity fields in such supersonic flow.

Statements and Declarations

No funding was received to assist with the preparation of this manuscript.

Acknowledgments

The authors thank Mr Carru, Cherubini, Roux, Taieb and Ms Lecointe for their assistance in conducting the experiments and optical diagnostics.

Yu, G. Li, J.G., Zhao, J.R., Yue, L.J., Chang, X.Y., Sung, C.-J. (2005) Proceedings of the Combustion Institute, 30:2859-2866